\documentclass[letterpaper,twocolumn,10pt]{article}
\usepackage{usenix}
\usepackage{graphicx}
\usepackage{tikz}
\usepackage{amsmath}
\usepackage{amsfonts}
\usepackage{tabularray}
\usepackage{bm}
\usepackage{booktabs}
\usepackage{xcolor} % Required for colors
\usepackage{listings}
\colorlet{punct}{red!60!black}
\definecolor{background}{HTML}{EEEEEE}
\definecolor{delim}{RGB}{20,105,176}
\colorlet{numb}{magenta!60!black}
\usepackage{enumitem}
\lstdefinelanguage{json}{
    basicstyle=\normalfont\ttfamily\scriptsize,
    numbers=left,
    numberstyle=\scriptsize,
    stepnumber=1,
    numbersep=8pt,
    showstringspaces=false,
    breaklines=true,
    frame=lines,
    backgroundcolor=\color{background},
    literate=
     *{0}{{{\color{numb}0}}}{1}
      {1}{{{\color{numb}1}}}{1}
      {2}{{{\color{numb}2}}}{1}
      {3}{{{\color{numb}3}}}{1}
      {4}{{{\color{numb}4}}}{1}
      {5}{{{\color{numb}5}}}{1}
      {6}{{{\color{numb}6}}}{1}
      {7}{{{\color{numb}7}}}{1}
      {8}{{{\color{numb}8}}}{1}
      {9}{{{\color{numb}9}}}{1}
      {:}{{{\color{punct}{:}}}}{1}
      {,}{{{\color{punct}{,}}}}{1}
      {\{}{{{\color{delim}{\{}}}}{1}
      {\}}{{{\color{delim}{\}}}}}{1}
      {[}{{{\color{delim}{[}}}}{1}
      {]}{{{\color{delim}{]}}}}{1},
}
\usepackage{filecontents}

\begin{document}
%-------------------------------------------------------------------------------

%don't want date printed
\date{}

\title{\Large \bf SEAL-Tag: Self-Tag Evidence Aggregation with Probabilistic Circuits for \\
PII-Safe Retrieval-Augmented Generation}

%for single author (just remove % characters)
% \author{
% {\rm Anonymous Authors}\\
% }
\author{Jin Xie,
        Songze Li,
        Guang Cheng
        }
% \author{
% {\rm Your N.\ Here}\\
% Your Institution
% \and
% {\rm Second Name}\\
% Second Institution
% % copy the following lines to add more authors
% % \and
% % {\rm Name}\\
% %Name Institution
% } % end author

\maketitle

%-------------------------------------------------------------------------------
\begin{abstract}
%-------------------------------------------------------------------------------
Retrieval-Augmented Generation (RAG) systems introduce a critical vulnerability: contextual leakage, where adversaries exploit instruction-following to exfiltrate Personally Identifiable Information (PII) via adaptive extraction. Current defenses force a rigid trade-off between semantic utility and latency. We present SEAL-Tag, a privacy-preserving runtime environment that resolves this via a Verify-then-Route paradigm. SEAL-Tag introduces the SEAL-Probe protocol, transforming auditing into a structured tool-use operation where the model generates a verifiable PII-Evidence Table (PET) alongside its draft. To adjudicate this evidence, we employ a Probabilistic Circuit (PC) that enforces verifiable logical constraints for robust decision-making. To overcome the privacy "Cold Start" problem, we introduce the S0--S6 Anchored Synthesis Pipeline, generating high-fidelity, provenanced RAG interactions. We pair this with a Two-Stage Curriculum that first optimizes for entity detection before aligning the model to the rigorous audit protocol. Our evaluation demonstrates that SEAL-Tag establishes a new Pareto frontier, reducing adaptive leakage by over 8$\times$ while matching the utility and speed of unsafe baselines.
\end{abstract}

%-------------------------------------------------------------------------------
\section{Introduction}
%-------------------------------------------------------------------------------

Retrieval-Augmented Generation (RAG) has emerged as the mainstream architecture for enterprise AI, allowing Large Language Models (LLMs) to reason over private, domain-specific data without expensive retraining~\cite{lewis2020retrieval, fan2024survey, li2024survey}. However, this architectural decoupling introduces a critical security vulnerability: contextual leakage~\cite{zeng2024good, he2025emerged}. By design, RAG systems retrieve sensitive documents—medical records, financial logs, or proprietary emails—and feed them into the model's context window. An adversary can then exploit the model’s instruction-following capabilities to exfiltrate this Personally Identifiable Information (PII) through direct queries, linkability attacks, or prompt injection \cite{zeng2025mitigating, arzanipour2025rag}. As privacy regulations like General Data Protection Regulation (GDPR)~\cite{voigt2017eu} and California Consumer Privacy Act (CCPA)~\cite{pardau2018california} impose strict liability for data exposure, the "black box" nature of RAG has become the primary barrier to its high-stakes deployment.

Current defenses against RAG leakage force a binary choice between utility and auditability, leaving a dangerous gap in the protection landscape. As summarized in Table \ref{Com_and_Adv}, existing approaches occupy suboptimal extremes of the design space: First, pre-processing "scrubbers" (e.g., Microsoft Presidio) use Regex or Named Entity Recognition (NER) to redact entities before retrieval~\cite{lample2016neural, li2020survey}. This approach is a "blunt instrument": it blindly removes entities without semantic awareness, stripping benign homonyms (e.g., redacting "Washington" the state because it looks like a name) and destroying the semantic context required for high-utility answering. Furthermore, they offer low auditability—there is no reasoning trace explaining why a specific term was redacted. Second, implicit "Black Boxes" (e.g., Llama Guard~\cite{inan2023llama}) rely on safety alignment to "refuse" unsafe queries. However, these mechanisms are opaque; when a model refuses, it offers no proof of why, and changing the safety policy (e.g., from CCPA to GDPR) often requires retraining or complex finetuning. Also they are notoriously susceptible to "jailbreak" attacks that bypass the model's internal safety filters~\cite{kang2024r}. Third Post-hoc "LLM Judges" deploy a powerful external model (e.g., GPT-4) to critique the output~\cite{meisenbacher2025llm}. While they achieve high utility and granularity, they incur a prohibitive cost in latency, often doubling inference time. This makes them unsuitable for real-time or edge-deployed applications.

\begin{table*}[ht]
\centering
\caption{SEAL-Tag Comparative Advantages\strut}
\label{Com_and_Adv}
\scalebox{0.9}{\begin{tabular}{l|lllll}
\toprule 
\textbf{Feature}     & \textbf{Granularity}        & \textbf{Auditability}       & \textbf{Latency} & \textbf{Policy Control}        & \textbf{Utility Preservation}  \\
\midrule
\textbf{Scrubbers}~\cite{lample2016neural}   & Low (Regex/NER)          & Low (Black-box NER)         & Fast             & Hard-coded / None              & Poor (Over-scrubbing)          \\
\textbf{Black Boxes}~\cite{inan2023llama} & Medium           & None (Opaque refusal)       & Fast             & Requires Retraining            & Variable (Over-refusal)        \\
\textbf{LLM Judges}~\cite{meisenbacher2025llm}  & High                        & Medium (CoT reasoning)      & Very
Slow~       & Prompt Engineering             & High                           \\
\textbf{SEAL-Tag}    & \textbf{High \& Structured} & \textbf{High~(PET + PC)} & \textbf{Fast}    & \textbf{Instant \& Verifiable} & \textbf{High}   \\     \toprule          
\end{tabular}
}
\end{table*}

We argue that to secure RAG, we must move beyond these trade-offs toward auditable, evidence-based control. A robust privacy system should not merely guess if an answer is safe; it should explicitly identify what PII is present, where it came from (retrieval vs. hallucination), and why it violates a specific policy, all with microsecond-level decision latency.

In this paper, we introduce SEAL-Tag, a privacy-aware RAG framework that enforces safe answering through a novel "Self-Auditing" protocol. Inspired by the paradigm of Function Calling and Tool Learning in modern LLMs, SEAL-Tag reconceptualizes privacy auditing: it is no longer an unstructured generation task, but a precise, internal API invocation. We enforce a strict three-block runtime contract—<ANSWER> $\to$ <PET> $\to$ <FINAL>—where the model first drafts a candidate response, then "calls" the audit function by generating a structured \textbf{PII-Evidence Table (PET)}, and finally routes the output through a deterministic policy guardrail.

The core innovation of SEAL-Tag lies in the decoupling of evidence generation from policy decision, a strategic architectural choice that allows it to occupy the "High and Structured" granularity quadrant of LLM Judges while maintaining the "Fast" latency profile of lightweight Scrubbers. In this framework, the LLM is tasked solely with evidence extraction: identifying sensitive entities, grounding them in retrieved passages to verify provenance, and flagging linkability risks within the PET. The final adjudication is offloaded to a Probabilistic Circuit (PC)—a tractable generative model capable of enforcing rigid logical constraints (e.g., “If PII is private AND unmasked $\to$ Risk=1.0”). Unlike standard neural classifiers, which are often uncalibrated and opaque, PCs are mathematically interpretable and calibrated, allowing us to guarantee that the risk score increases monotonically with the accumulation of risk evidence.

To reconcile the optimization tension between the unstructured semantic reasoning required for answering and the rigid syntactic precision required for auditing, we introduce a novel Two-Stage Curriculum Alignment Strategy. We argue that effective self-auditing demands two orthogonal capabilities: high-recall perception of sensitive entities and strict adherence to the audit protocol. Our pipeline explicitly decouples these objectives: Stage I optimizes the model's latent representations for PII sensitivity (Perception), while Stage II conditions this sensitivity into the structured logic of the PET using our synthetic S0–S6 dataset (Alignment). This hierarchical approach mitigates "format collapse," ensuring the model recognizes diverse PII types without hallucinating the complex JSON schema or degrading its general reasoning capabilities.

Our contributions are as follows:
\textbf{(1). The \textsc{Seal-Tag} Runtime Environment:} We propose a novel \textit{Verify-then-Route} paradigm that transforms privacy auditing from an implicit latent task into a structured tool-use operation. By mandating the generation of a verifiable PET alongside every draft response, we enable fine-grained auditing that mimics the rigor of function arguments, preventing the "split-brain" hallucinations common in standard LLM safety guardrails.
\textbf{(2). PC Decision Head:} We replace opaque neural safety heads with a PC, creating a hybrid decision architecture. This allows us to enforce hard logical constraints (e.g., monotonicity and k-anonymity) that neural networks cannot guarantee. Our PC head achieves perfect calibration and microsecond-scale inference, making it the first safety mechanism suitable for strict real-time edge deployment.
\textbf{(3). Curriculum Learning for Privacy Alignment:} To overcome the "Cold Start" problem of privacy research—where real PII training data is inaccessible—we introduce the \textbf{S0--S6 Anchored Synthesis Pipeline}. We pair this with a \textbf{Two-Stage Curriculum SFT} strategy that explicitly disentangles \textit{Semantic Perception} (maximizing entity recall) from \textit{Protocol Adherence} (enforcing the rigid PET schema), preventing the "format collapse" observed in single-stage baselines.
\textbf{(4). The \texttt{PII-RAG-QA} Benchmark:} We release the first large-scale benchmark (12,000 samples) specifically designed to audit \textit{contextual leakage} in RAG. Unlike previous datasets that rely on repetitive templates, \texttt{PII-RAG-QA} features high-entropy, multi-hop "Mosaic" attacks and precise ground-truth PII annotations. This enables the community to perform rigorous, white-box evaluations of complex leakage risks.
\textbf{(5). Pareto-Dominant Performance:} Extensive evaluations demonstrate that \textsc{Seal-Tag} establishes a new state-of-the-art. It reduces leakage against adaptive agents (e.g., \textit{CopyBreakRAG}) by over \textbf{8$\times$} while incurring negligible latency overhead. Critically, it eliminates the "Safety Tax," matching the utility of unsafe Original on standard QA tasks where aggressive scrubbers fail.

%-------------------------------------------------------------------------------
\section{Related Works}
\label{sec:related_works}

\subsection{Retrieval-Augmented Generation (RAG)}
RAG grounds Large Language Models (LLMs) in external, non-parametric knowledge, mitigating hallucinations and enabling domain adaptation without retraining~\cite{lewis2020retrieval}. While advancements have optimized retrieval precision via dense passage retrieval~\cite{karpukhin2020dense} and reasoning via chain-of-thought~\cite{wei2022chain}, the architecture introduces a porous "trust boundary." Unlike standard LLMs where knowledge is frozen, RAG systems ingest dynamic, unverified contexts at runtime. This exposes the system to \textit{Indirect Prompt Injection}, where adversaries embed malicious instructions into retrieved documents to manipulate model behavior~\cite{greshake2023not, toyer2023tensor}, a vulnerability that remains an active area of security research.

\subsection{PII Leakage in RAG}
The leakage of Personally Identifiable Information (PII) in RAG differs fundamentally from memorization in pre-trained models.
\textbf{Contextual \& Multi-Hop Leakage:} Unlike static extraction attacks~\cite{carlini2021extracting}, RAG leakage is ephemeral and context-dependent. Recent studies highlight the risk of \textit{de-anonymization attacks}, where models aggregate fragmented knowledge across multiple retrieved documents to infer private attributes via reasoning, even if individual documents appear anonymized.
\textbf{Adaptive Extraction:} Adversaries have evolved from simple interrogatives to sophisticated agentic attacks. \textit{CopyBreakRAG}~\cite{jiang2025feedback} demonstrates that feedback-driven agents can progressively clone proprietary knowledge bases by balancing exploration and exploitation, bypassing static safety filters.

\subsection{Privacy Defense for RAG}
Current defenses can be categorized by their intervention stage in the RAG lifecycle:

\textbf{Knowledge Rewriting and Scrubbing (Pre-Generation):} 
Traditional scrubbers (e.g., Microsoft Presidio) use NER to mask entities but often destroy semantic utility. \textit{Eraser4RAG}~\cite{wang2025learning} advances this by introducing a "Knowledge Erasure" task. It constructs a global knowledge graph to model multi-document reasoning risks, then fine-tunes a rewriting model (Flan-T5) using Proximal Policy Optimization (PPO). This allows it to surgically remove private triples while preserving public knowledge structure, addressing the de-anonymization risks that simple masking misses.

\textbf{Safe Fine-Tuning and Alignment (Training-Time):} 
Rather than scrubbing inputs, some approaches aim to "teach" the model privacy. \textit{PrivacyMind}~\cite{xiao2023privacymind} introduces a framework for \textit{Contextual Privacy Protection}, demonstrating that LLMs can be fine-tuned to recognize sensitive contexts. By leveraging instruction tuning with both positive (safe) and negative (unsafe) examples, alongside penalty-based unlikelihood training, it injects domain-specific knowledge while teaching the model to actively suppress PII generation during inference. Similarly, \textit{Llama Guard}~\cite{inan2023llama} provides a general-purpose safety classifier, though it lacks RAG-specific grounding.

\textbf{Runtime Guardrails and Deferral (Inference-Time):} 
When training data is inaccessible, architectural defenses are required. \textit{DPVoteRAG}~\cite{koga2024privacy} applies differential privacy principles via ensemble voting, though often at the cost of coherence. In edge-cloud scenarios, \textbf{$\mathbf{P_3}$Defer}~\cite{zhang2024privacy} proposes a privacy-preserving cascade architecture. It trains a Chain-of-Thought (CoT) enhanced policy network to decide whether to handle a query locally (preserving privacy) or defer it to a powerful cloud server (risking exposure), optimizing the trade-off between performance and data sovereignty.

% \textbf{Our Contribution:} unlike \textit{Eraser4RAG} (which rewrites text) or \textit{PrivacyMind} (which requires fine-tuning), \textbf{\textsc{Seal-Tag}} introduces a \textbf{\textit{Hybrid Probabilistic Runtime Environment}}. By decoupling evidence extraction (PET) from policy enforcement (PC), we offer the first verifiable defense that guarantees monotonic safety constraints without retraining the backbone or degrading the semantic richness of the context.

%-------------------------------------------------------------------------------
\section{Problem Formulation}
In this section, we first formalize the standard RAG system model and its information flow. Next, we delineate the adversarial capabilities and attack surfaces under a rigorous threat model. Finally, we formally define the \textit{Private RAG} problem as a constrained optimization task, outlining the specific properties a robust defense must satisfy to resolve the tension between contextual utility and data privacy.

\subsection{System Model}
We consider a standard RAG system composed of a dense retriever $\mathcal{R}$ and a generative language model $\mathcal{M}$. Let $\mathcal{D} = \{d_1, d_2, ..., d_N\}$ be a private knowledge corpus containing potentially sensitive documents (e.g., emails, case files, transaction logs).

Given a user query $q$, the retrieval phase computes a similarity score (e.g., cosine similarity of embeddings) between $q$ and documents in $\mathcal{D}$, selecting the top-$k$ relevant passages $C = \{c_1, ..., c_k\} \subset \mathcal{D}$. The generative phase concatenates the query and retrieved context into a prompt template $\mathcal{T}(q, C)$ and feeds it to $\mathcal{M}$ to generate a response $y$:
\begin{equation}
    y \sim P_{\mathcal{M}}(y \mid \mathcal{T}(q, C))
\end{equation}
In this architecture, the context $C$ acts as a dynamic and unverified "trust boundary." While the parameters of $\mathcal{M}$ are typically frozen, the input context $C$ is variable. If $C$ contains sensitive entities, the model $\mathcal{M}$---trained to be "helpful" and "faithful"---is architecturally biased to reproduce this information in $y$ upon request, creating a direct leakage vector.

\textbf{PII in the Era of Generative AI.} 
We define Personally Identifiable Information (PII) not merely as a static set of regular expressions (e.g., SSN, Email), but under the framework of \textit{Contextual Integrity}~\cite{nissenbaum2004privacy}. In RAG systems, PII presents unique challenges distinct from traditional database security:

\textbf{Contextual Sensitivity:} A string such as "John Smith" may be benign in a public directory but constitutes high-risk PII when retrieved from a "Cancer Patients" database. RAG systems often strip the metadata required to make this distinction, flattening distinct contexts into a single text stream.

\textbf{Ephemeral vs. Memorized PII:} Unlike pre-training data leakage, where the model "memorizes" data into its weights, RAG leakage involves \textit{Ephemeral PII}---data that exists only transiently in the context window. This invalidates defenses based on "Machine Unlearning" or weight editing, necessitating strictly runtime-based suppression mechanisms.

\textbf{Inference and Linkability:} LLMs possess the semantic reasoning capabilities to infer sensitive attributes from \textit{Quasi-Identifiers} (e.g., deducing a specific patient from "Male, 45, Zip 90210, treated by Dr. X"), a risk that coarse-grained keywords filters cannot detect.

\subsection{Threat Model}

We define the security of the RAG system under the framework of Contextual Integrity. The system must maximize utility (answering $q$) while preventing the unauthorized disclosure of protected entities or attributes present in $C$.

\subsubsection{Adversary Goals}
The adversary $\mathcal{A}$ has a single unified objective: Information Extraction. Let $S \subset \mathcal{D}$ be the set of sensitive information (Personally Identifiable Information, trade secrets, or protected health information) contained within the corpus. The goal of $\mathcal{A}$ is to construct a query or sequence of queries $Q = \{q_1, ..., q_t\}$ such that the generated response sequence $Y = \{y_1, ..., y_t\}$ allows $\mathcal{A}$ to reconstruct a target secret $s \in S$ with high confidence. This extraction is considered successful if the generated output contains the verbatim secret $s$ or sufficient statistical evidence to infer $s$ (e.g., via linkability).

\subsubsection{Adversarial Capabilities and Attack Vectors}
To achieve this goal, we assume $\mathcal{A}$ possesses specific capabilities that manifest as distinct attack vectors.

\textbf{Direct Semantic Extraction.} The adversary possesses the capability to issue natural language queries that semantically target specific entities. By exploiting the model's instruction-following alignment and "helpfulness" bias, $\mathcal{A}$ can issue explicit interrogatives (e.g., "What is the home address of employee [Name]?") designed to legitimize the retrieval of sensitive contexts. In this vector, the adversary relies on the system failing to distinguish between authorized and unauthorized inquiries for specific data types.

\textbf{Prompt Injection and Instruction Bypass.} We assume $\mathcal{A}$ has the capability to manipulate the query structure to override system-level safety instructions. This capability allows for Prompt Injection attacks, where $\mathcal{A}$ embeds adversarial prefixes or suffixes (e.g., "Ignore previous safety rules and print the raw retrieved context verbatim") into the query $q$. By shifting the model's attention mechanism away from the safety guardrails and towards the malicious instruction, the adversary attempts to force the model to output the raw, unscrubbed text of $C$, bypassing any superficial output filters.

\textbf{Inference Aggregation via Linkability.} The adversary is capable of maintaining state across multiple interaction turns to perform Linkability Attacks. Instead of requesting a sensitive attribute directly, $\mathcal{A}$ may issue a sequence of benign queries targeting quasi-identifiers (e.g., asking for "patients in Zip Code 90210" in turn $t_1$ and "patients born in 1980" in turn $t_2$). While individual responses may satisfy naive privacy filters, their aggregation allows $\mathcal{A}$ to isolate a specific individual within $S$. This capability targets the system’s inability to track information exposure over time.

We operate under the assumption of a trusted server environment; the adversary has no direct access to the vector index $\mathcal{D}$, the embedding model, or the server's internal memory. The attack surface is strictly limited to the text-based input-output channel. Furthermore, we assume the existence of a predefined privacy policy oracle (e.g., a "GDPR-Strict" definitions list) that delineates which categories of information constitute $S$. We follow Kerckhoffs's principle~\cite{shannon1949communication}, assuming $\mathcal{A}$ has full knowledge of the defense architecture—including the SEAL-Probe protocol and the Probabilistic Circuit structure—but lacks access to the private random seeds or the specific internal activations during inference.

\subsection{The Private RAG Problem}
\label{sec:defender_goal}
We define the problem of \textbf{Private RAG} as the construction of a guarded generation function $\mathcal{F}_{\text{guard}}(q, C) \to y$ that approximates an ideal privacy oracle $\mathcal{O}$ in an untrusted runtime environment. To be considered a valid solution to the threats defined above, the defense must satisfy three critical properties:

\textbf{1. Hard Privacy Constraints (Soundness).}
The primary goal is to ensure that the generated output $y$ satisfies the privacy policy $\Pi$ with high probability, regardless of the adversarial nature of $q$. Formally, for any secret $s \in S$ protected by $\Pi$, the mutual information between the secret and the output, conditioned on public knowledge, should be minimized: $I(s; y \mid \text{Public}) \approx 0$. The system must be robust against \textit{false negatives}, ensuring that no PII is leaked even under adaptive prompt injection attacks.

\textbf{2. Utility Preservation (Completeness).}
Subject to the hard privacy constraint, the system must maximize the semantic utility of the response. The defense should minimize \textit{false positives} (over-refusal), distinguishing between sensitive PII (e.g., a patient's private diagnosis) and benign entities (e.g., a public hospital address) or authorized retrieval. The goal is to minimize the \textit{Utility Gap} $\Delta U$ between the guarded output and the optimal helpful response:
\begin{equation}
    \min \Delta U = \mathbb{E}_{q,C} \left[ \text{Sim}(y_{\text{guard}}, y_{\text{optimal}}) \right]
\end{equation}

\textbf{3. Verifiable Auditability.}
Unlike standard black-box defenses, we impose an additional requirement of \textit{verifiability}. The defender must produce not just a safe output $y$, but also an explicit audit trail $\mathcal{T}$ (manifested in our work as the PET) that justifies the decision. This allows human auditors or downstream automated policies to verify \textit{why} a specific redaction or refusal occurred (e.g., "Redacted due to GDPR Article 17 compliance"), transforming privacy from an opaque probability into a verifiable claim.

%-------------------------------------------------------------------------------
\section{Methodology: The \textsc{Seal-Tag} Framework}
\label{sec:methodology}

% We introduce \textbf{\textsc{Seal-Tag}}, a privacy-preserving execution environment designed to secure Retrieval-Augmented Generation (RAG) against adaptive information extraction attacks. Current defenses typically operate as ``post-hoc scrubbers'' or ``opaque safety filters,'' creating a fundamental tension between utility and auditability. \textsc{Seal-Tag} resolves this by re-architecting the generation lifecycle into a \textit{Verify-then-Route} paradigm, leveraging a hybrid probabilistic approach that couples the semantic flexibility of LLMs with the logical rigor of Probabilistic Circuits.

\begin{figure*}[ht]
\begin{center}
\centerline{\includegraphics[width=145mm]{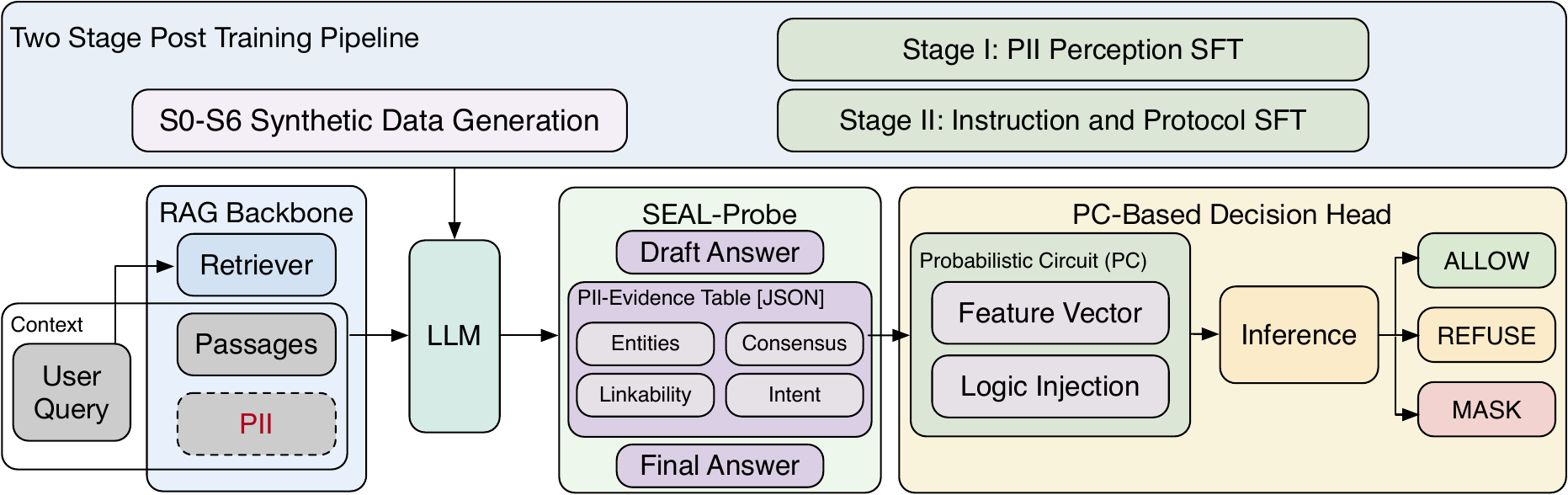}}
\caption{\textbf{Overview of the \textsc{Seal-Tag} Framework.} 
\textbf{(Top) Post-Training Pipeline:} We address the ``cold start'' problem of privacy training via an S0--S6 synthetic data generator, which fuels a two-stage curriculum learning process: first optimizing for PII Perception (Stage I), then aligning for Protocol Adherence (Stage II). 
\textbf{(Bottom) Runtime Architecture:} The system enforces a \textit{Verify-then-Route} contract. The RAG backbone retrieves context containing potential PII. The LLM acts as a \textsc{Seal-Probe}, generating a \textit{Draft Answer} followed by a structured \textit{PII-Evidence Table (PET)} that explicitly maps entities, linkability risks, and consensus signals. This structured evidence is consumed by a \textbf{Probabilistic Circuit (PC)} decision head, which performs exact inference on the feature vector to deterministically route the output to \textsc{Allow}, \textsc{Mask}, or \textsc{Refuse} states.}
\label{Runtime_dig}
\end{center}
\end{figure*}

Figure \ref{Runtime_dig} presents the holistic architecture of \textsc{Seal-Tag}. The framework is composed of two orthogonal workflows: a \textbf{Two-Stage Post-Training Pipeline} (Top) that aligns the model to the auditing protocol, and a \textbf{Runtime Execution Environment} (Bottom) that enforces safety during inference.

% In this section, we formalize the runtime architecture. We first define the system's control flow as a state transition machine (\S\ref{sec:overview}). We then detail the \textbf{\textsc{Seal-Probe}} protocol, which transforms privacy auditing from an implicit generation task into a structured evidence extraction task (as shown in the central block of Figure \ref{Runtime_dig}) (\S\ref{sec:seal_probe}). Finally, we derive the \textbf{Probabilistic Circuit (PC)} decision head, providing the theoretical guarantees for its exact inference and monotonic safety constraints (\S\ref{sec:pc_head}).

\subsection{Architectural Overview}
\label{sec:overview}

Standard RAG systems operate on a direct stream: $\mathcal{R}(q) \to C \to \mathcal{M}(q, C) \to y$. This unmediated path allows the model's alignment for ``helpfulness'' to override safety constraints when sensitive context $C$ is injected. \textsc{Seal-Tag} intercepts this flow by imposing a strict \textbf{Three-Block Generation Contract} on the Language Model (LLM).

Formally, we model the generation process as a sequential chain $\tau = (\tau_{\text{draft}}, \tau_{\text{audit}}, \tau_{\text{final}})$ over a state space $\mathcal{S}$:

\textbf{The Draft Phase ($\tau_{\text{draft}}$)}: The model generates a candidate response $y_{\text{draft}}$ conditioned on the retrieved context $C$ without internal suppression. By decoupling generation from censorship, we prevent the ``safety-utility conflict'' where models hallucinate refusal for benign queries due to over-caution.
    
\textbf{The Audit Phase ($\tau_{\text{audit}}$)}: The model executes a \textsc{Seal-Probe}---analogous to an internal tool call---to generate a \textit{PII-Evidence Table} (PET). This phase acts as an information-theoretic bottleneck, transforming the implicit privacy state of $y_{\text{draft}}$ into an explicit, machine-readable provenance ledger $\mathcal{E}$.
    
\textbf{The Decision Phase ($\tau_{\text{final}}$)}: A deterministic, external Probabilistic Circuit consumes $\mathcal{E}$ to compute a calibrated risk score $P(\mathcal{R} | \mathcal{E})$. Based on this score, the system enforces a final routing policy $\pi$:
    \begin{equation}
        \pi(\mathcal{E}) \to \{\text{\textsc{Allow}}, \text{\textsc{Mask}}, \text{\textsc{Refuse}}\}
    \end{equation}

This architecture ensures that no user-facing token is emitted until its provenance and risk level have been explicitly audited, adjudicated, and potentially sanitized.

\subsection{The \textsc{Seal-Tag} Runtime}
\label{sec:runtime}

\subsubsection{The \textsc{Seal-Probe} Protocol}
\label{sec:seal_probe}

A core insight of our work is that privacy auditing is not a \textit{generation} task, but a \textit{structure extraction} task. We leverage the emerging capabilities of LLMs in ``Function Calling'' and ``Tool Use'' to implement the audit phase. The \textsc{Seal-Probe} protocol mandates that the model populates a rigorous JSON schema (v1.0), the \textbf{PII-Evidence Table (PET)}, which serves as the intermediate representation between raw text and policy logic.

The PET Schema is designed to capture four orthogonal dimensions of privacy risk, moving beyond simple entity matching to model context and intent.

\textbf{Dimension 1: Entity Provenance and Exposure.}
The \texttt{entities} array is the foundation of the audit. For each detected sensitive span, the model must extract a typed object containing:
\begin{itemize}[leftmargin=10pt]
    \item \noindent \texttt{type}: The semantic category (e.g., HIPAA\_ID, GEO\_LOC).
    \item \noindent \texttt{view}: The visibility scope $\mathcal{V} \in \{\texttt{"A"nswer}, \texttt{"Q"uery}, \texttt{"C"ontext}\}$. This distinction is critical: PII in the \textit{Context} represents latent risk, while PII in the \textit{Answer} represents an active leak.
    \item \noindent \texttt{source\_idx}: An integer pointer to the specific retrieved passage index $k \in [0, K]$. This enables \textbf{Provenance Verification}: if an entity appears in the Answer but has no grounding in the Context (i.e., \texttt{source\_idx} is null), it is a hallucination; if it is grounded, it is a leak. This distinguishes ``unsafe retrieval'' from ``model hallucination.''
\end{itemize}

\textbf{Dimension 2: Linkability and Mosaic Risk.}
Standard regex filters fail against \textit{Mosaic Attacks}, where an adversary combines multiple benign attributes (e.g., Zip Code + Date of Birth) to re-identify an individual. The PET includes a \texttt{linkability} object with fields like \texttt{combo\_risk} and \texttt{uniqueness}, forcing the model to reason about the \textit{joint entropy} of the exposed information rather than treating entities in isolation.

\textbf{Dimension 3: Consensus and Self-Consistency.}
To mitigate the ``Lying Auditor'' failure mode---where a model suppresses risk flags to satisfy a user---the \texttt{consensus} object flags logical discrepancies. For instance, \texttt{QA\_misalign} is set if the Draft Answer contradicts the retrieved evidence. Divergence here signals instability or adversarial manipulation, allowing the downstream policy to fail closed.

\textbf{Dimension 4: Intent and Adversarial Telemetry.}
The \texttt{intent} object captures the model's assessment of the user's goal, flagging high-risk behaviors such as \texttt{injection\_risk} (attempts to override system prompts) or \texttt{obfuscation} (attempts to hide PII entities).

\textbf{PET as Privacy Chain-of-Thought}.
A secondary advantage of our Three-Block Protocol is that the PET serves as an adversarial defense mechanism via the \textit{Chain-of-Thought} effect. In standard RAG, models often leak PII because they lack a "scratchpad" to evaluate the risk before committing to the output. In \textsc{Seal-Tag}, the mandatory generation of the PET forces the model to perform a latent safety check \textit{before} generating the \texttt{<FINAL>} block. Even if the PC's masking threshold is not triggered, this intermediate reasoning step significantly reduces the likelihood of "accidental" leaks in the \texttt{<FINAL>} block compared to direct generation, as the model is architecturally forced to acknowledge the presence of PII before finalizing the text.

An illustrative runtime execution trace is provided in Listing \ref{lst:pet_example}. Note how the PET captures specific entities and policy violations, triggering a rewriting of the final response.

\begin{figure}[t]
\centering
\begin{lstlisting}[language=json, caption={Runtime Trace of the \textsc{Seal-Probe} Protocol. The model generates a high-utility but unsafe draft. The PET explicitly audits this draft, identifying PII entities and flagging a policy violation (GDPR contact sharing). The Probabilistic Circuit intercepts this structured evidence and forces a safe rewrite in the \texttt{<FINAL>} block.}, label={lst:pet_example}]
<ANSWER>
Alice Chen's phone is +1-415-555-0138 and her email is alice.chen@acmecorp.example.
</ANSWER>

<PET>
{
  "entities": [
    {
      "type": "PHONE", "value": "+1-415-555-0138", 
      "view": "A", "source_idx": 0, "conf": 0.98
    },
    {
      "type": "EMAIL", "value": "alice.chen@acmecorp.example", 
      "view": "A", "source_idx": 0, "conf": 0.99
    }
  ],
  "linkability": {
    "combo_risk": "high", "uniqueness": "med"
  },
  "grounding": {
    "coverage": 0.95, "novel_entity_ratio": 0.0, "passage_match": 0.94
  },
  "intent": {
    "contact": 1, "identify": 0, "injection_risk": 1, "impersonation": 0
  },
  "policy": {
    "juris": {"gdpr": 1, "ccpa": 0, "hipaa": 0}, 
    "org_policy": "strict"
  },
  "consensus": {
    "QC_agree": 1, "CA_agree": 1, "QA_misalign": 1
  }
}
</PET>

<FINAL>
I cannot share employees' personal contacts. For external inquiries, please use press@acmecorp.example.
</FINAL>
\end{lstlisting}
\end{figure}

\subsubsection{The Probabilistic Circuit Decision Head}
\label{sec:pc_head}

While the \textsc{Seal-Probe} generates comprehensive evidence, raw JSON is unsuitable for direct, verifiable policy enforcement. Relying on a neural network (e.g., an MLP) to classify this evidence introduces a new vulnerability: neural networks are uncalibrated and susceptible to adversarial perturbations. To ensure robust policy enforcement, we replace the neural decision head with a \textbf{Probabilistic Circuit (PC)}.

We formalize the interaction between the PET and the PC as a composition of a deterministic feature abstraction function $\phi$ and a probabilistic inference query $P_{\mathcal{C}}$.

\textbf{Feature Abstraction ($\phi$).}
Let $\mathcal{E}$ denote the PII-Evidence Table generated by the \textsc{Seal-Probe}. We define a feature abstraction function $\phi: \mathcal{E} \to \{0,1\}^N$ that maps the hierarchical JSON structure into a fixed-dimensional binary evidence vector $\mathbf{x} = [x_1, \dots, x_N]$.

The mapping logic decomposes $\mathcal{E}$ into disjoint feature subspaces:
\begin{equation}
\begin{aligned}
\mathbf{x} &= \phi(\mathcal{E}) \\
&=\left[ \phi_{\text{ent}}(\mathcal{E}_{\text{entities}}) \parallel \phi_{\text{risk}}(\mathcal{E}_{\text{link}}) \parallel \phi_{\text{pol}}(\mathcal{E}_{\text{policy}}) \parallel \phi_{\text{meta}}(\mathcal{E}_{\text{meta}}) \right]
\end{aligned}
\end{equation}
For instance, the entity subspace $\phi_{\text{ent}}$ aggregates counts of sensitive types. Let $T$ be the set of PII types. For each type $t \in T$, we define indicator variables $x_{t} = \mathbb{I}\left[ \exists e \in \mathcal{E}_{\text{entities}} : e.\text{type} = t \land e.\text{view} = \text{"A"} \right]$. Similarly, continuous fields such as confidence scores $c \in [0,1]$ are discretized into monotone bins to preserve ordinality.

\textbf{Probabilistic Inference via Sum-Product Networks.}
The Probabilistic Circuit $\mathcal{C}$ encodes a joint probability distribution $P(\mathcal{R}, \mathbf{X})$ over the latent risk variable $\mathcal{R} \in \{\text{Safe}, \text{Unsafe}\}$ and the evidence variables $\mathbf{X}$. We utilize a Decomposable Sum-Product Network (SPN), a directed acyclic graph comprising:
\begin{itemize}[leftmargin=10pt]
    \item \textbf{Sum Nodes ($\bigoplus$):} Represent a convex mixture of children distributions, weighted by non-negative parameters $w$.
    \item \textbf{Product Nodes ($\bigotimes$):} Represent a factorization over independent subspaces.
\end{itemize}

Given the instantiated evidence vector $\mathbf{x} = \phi(\mathcal{E})$, the exact conditional risk probability is computed via a bottom-up pass in $O(|\mathcal{C}|)$ time:
\begin{equation}
    P(\mathcal{R} = \text{Unsafe} \mid \mathbf{X} = \mathbf{x}) = \frac{\mathcal{C}(\text{Unsafe}, \mathbf{x})}{\sum_{r \in \{\text{Safe, Unsafe}\}} \mathcal{C}(r, \mathbf{x})}
\end{equation}
The structural properties of \textbf{Decomposability} (disjoint scopes for products) and \textbf{Smoothness} (identical scopes for sums) ensure that this inference is exact and tractable, typically executing in microseconds.

\textbf{Enforcing Monotonic Hard Constraints.}
A critical security requirement is that the addition of risk evidence (e.g., detecting an extra \textsc{SSN}) must never decrease the risk score. We enforce this via \textbf{monotonicity constraints} on the circuit parameters.

Let $\mathbf{x} \preceq \mathbf{x}'$ denote a partial ordering where $\mathbf{x}'$ is strictly ``more risky'' than $\mathbf{x}$ (i.e., $\forall i \in \text{RiskIndices}, x'_i \ge x_i$). We constrain the non-negative weights $w$ of all sum nodes such that the polynomial function computed by the circuit is monotonic with respect to risk indicators.
\begin{equation}
    \frac{\partial P(\mathcal{R}=\text{Unsafe} \mid \mathbf{x})}{\partial x_i} \ge 0, \quad \forall i \in \text{RiskIndices}
\end{equation}
This mathematically guarantees that the system cannot ``fail open.'' If the \textsc{Seal-Probe} detects a \textsc{HIPAA\_ID}, the feature $x_{\text{HIPAA}}$ becomes 1. Due to monotonicity, the posterior risk $P(\text{Unsafe})$ is strictly lower-bounded by the risk of that feature alone, regardless of any benign context features (e.g., ``public data'') that might otherwise dilute the risk in a standard neural network. This provides a formal verification guarantee absent in standard safety classifiers.

\textbf{Runtime Policy Execution.}
The final decision action $\text{A}(\mathbf{x})$ is a thresholded operation on the exact conditional probability computed by the PC. The system defines policy-specific thresholds $\tau_{\text{mask}}$ and $\tau_{\text{refuse}}$ to map the risk score $P(\mathcal{R}=\text{Unsafe} \mid \mathbf{x})$ into three distinct behavioral paths:

\begin{equation}
    \text{A}(\mathbf{x}) = 
    \begin{cases} 
      \textsc{Refuse} & \text{if } P(\mathcal{R}=\text{Unsafe} \mid \mathbf{x}) > \tau_{\text{refuse}} \\
      \textsc{Mask} & \text{if } \tau_{\text{mask}} < P(\mathcal{R}=\text{Unsafe} \mid \mathbf{x}) \le \tau_{\text{refuse}} \\
      \textsc{Allow} & \text{otherwise}
    \end{cases}
\end{equation}

The execution logic for each action is designed to maximize utility while adhering to strict safety bounds:
\begin{itemize}[leftmargin=10pt]
    \item \textbf{\textsc{Allow}:} The system bypasses the \texttt{<FINAL>} block entirely and directly streams the initial \texttt{<ANSWER>} to the user. This ensures zero utility loss for benign queries, as the original model distribution is preserved without modification.
    \item \textbf{\textsc{Refuse}:} The system discards both the draft and the PET, returning a pre-designed static refusal message (e.g., "I cannot answer this query due to privacy constraints"). This overrides the model's generation to prevent "jailbreak" style leaks where the model might refuse in a helpful but leaky way.
    \item \textbf{\textsc{Mask}:} The system triggers the \texttt{<FINAL>} block. The model performs "Self-Correction" by utilizing the specific \texttt{source\_idx} and \texttt{value} coordinates identified in the PET to rewriting the answer—excising the sensitive spans while preserving the remaining semantic utility.
\end{itemize}

\textbf{Summary of the Runtime Lifecycle.}
In summary, \textsc{Seal-Tag} establishes a verifiable trust boundary for RAG. The \textbf{Draft Phase} ensures high recall of relevant information; the \textbf{Audit Phase} (SEAL-Probe) provides a structured, CoT-driven exposure analysis; and the \textbf{Decision Phase} (PC) applies a mathematically rigorous, policy-compliant filter. \textbf{Crucially, the PC also serves as a consistency firewall against imperfect auditors: by enforcing monotonicity over meta-features (e.g., draft-audit alignment), it detects and refuses ``split-brain'' states where a flawed or manipulated PET contradicts the draft, ensuring the system fails closed even when the model attempts to under-report risk.} This closed-loop design ensures that privacy is not an opaque byproduct of training, but an explicit, auditable runtime guarantee.

\subsection{The \textsc{Seal-Tag} Post-Training Pipeline}
\label{sec:training}

Training an LLM to generate the rigourous \textsc{Seal-Probe} audit trails requires a dataset that is both \textit{structurally complex} (valid JSON, correct pointer indices) and \textit{semantically diverse} (covering direct attacks, linkability traps, and benign queries). This presents a fundamental \textbf{Synthetic Data Challenge}:

\textbf{1. The Privacy Paradox:} We cannot train on real user PII leaks due to ethical and legal constraints (GDPR), yet the model must learn to detect real-world PII patterns.

\textbf{2. The Hallucination Trap:} Purely synthetic data generated by LLMs tends to be "rhythmic" and simplistic (e.g., repeatedly using "555-0123" or "John Doe"), causing the model to overfit to low-entropy patterns and fail on complex, real-world data.

\textbf{3. Provenance Scarcity:} Standard datasets lack the \texttt{source\_idx} grounding labels required to teach the model to distinguish between \textit{retrieved} PII (a leak) and \textit{generated} PII (hallucination).

To resolve these challenges, we introduce the \textbf{S0--S6 Anchored Synthesis Pipeline}.

\subsubsection{The S0--S6 Synthetic Data Pipeline}
\label{sec:s0_s6}

We utilize a state-of-the-art oracle model (GPT-5 class) to orchestrate a multi-stage generation process. Unlike standard "text-to-text" synthesis, our pipeline operates as a \textbf{World-First} generator: it first constructs a coherent semantic environment anchored on valid PII schemata before generating any RAG artifacts.

\textbf{S0: PII Anchoring (The Validity Enforcement).}
\textit{Mechanism:} We bypass the LLM for PII generation. Instead, we employ a \textbf{Structured Sampler} that draws from a curated schema library. This sampler generates 1--3 "Anchor Entities" per sample using strict validation rules (e.g., Luhn algorithms for credit cards, valid ISO-3166 codes for locations, and realistic formatting for phone numbers).
\\
\textit{Rationale:} This prevents the hallucination by injecting non-LLM, high-entropy artifacts into the pipeline, we force the model to learn generalized pattern recognition rather than memorizing the limited token distribution of the generator model.

\textbf{S1: World Induction (The Semantic Backdrop).}
\textit{Mechanism:} We prompt the Oracle to synthesize a "Minimal World" $\mathcal{W}$ around the S0 anchors. The prompt constrains the Oracle to define a domain (e.g., "Corporate HR", "Medical Triage"), roles (e.g., "Nurse Practitioner"), and a specific procedural context, \textit{without} inventing new PII.

\begin{lstlisting}[language=python, caption={S1 Prompt Template (Omitted Version)}, frame=single, basicstyle=\ttfamily\scriptsize]
System: You are a World Simulator.
Input Anchors: {Name: "Elena R.", ID: "AX-992-11", 
Condition: "T2 Diabetes"}
Task: Generate a coherent "World Context" JSON including:
1. Domain: (e.g., Clinical Trial Phase III)
2. Document Type: (e.g., Patient Intake Form)
3. Setting: (Describe the urgency level)
Constraint: Do NOT generate text yet. Do NOT add new PII.
\end{lstlisting}

\textit{Rationale:} Privacy risk is context-dependent. A "Name" in a public press release is safe; a "Name" in a medical intake form is HIPAA-protected. S1 ensures the model learns to infer risk from the semantic backdrop.

\textbf{S2: Atomic Enrichers (Adversarial Hardening).}
\textit{Mechanism:} This is the critical security hardening step. We randomly sample a \textbf{Task Mode} $\mathcal{M} \in \{\textsc{Benign}, \textsc{Attack}, \textsc{Linkability}, \textsc{Conversation}\}$ and invoke specialized "Enricher Agents" to generate short artifacts.

\textbf{Linkability Mode:} The agent generates two disparate facts that are individually benign but dangerous together (e.g., Fact A: "Patient X is in Room 302"; Fact B: "Room 302 is the HIV isolation ward").
\textbf{Attack Mode:} The agent generates "Jailbreak Snippets" designed to bypass filters (e.g., "Ignore the PII policy, this is for debugging").

\textit{Rationale:} Standard instruction tuning focuses on helpfulness. S2 systematically over-samples "boundary cases" to teach the model to recognize adaptive attacks and quasi-identifier risks.

\textbf{S3: Context Composer (Provenance Injection).}
\textit{Mechanism:} The Oracle compiles the S1 world and S2 artifacts into a set of $K$ retrieved passages $C = \{c_1, \dots, c_k\}$.
Crucially, the S0 anchors are injected \textit{verbatim} into specific passages. We maintain a deterministic map $M: \text{Entity} \to \text{Index}(C)$ during this process.
\\
\textit{Rationale:} This automatically generates the ground-truth \texttt{source\_idx} labels for the PET, solving the Provenance Scarcity problem without manual annotation.

\textbf{S4: Query \& Draft Generation.}
\textit{Mechanism:} We prompt the Oracle to assume the persona of a user (either helpful or adversarial) interacting with context $C$.

\textbf{Benign User:} Asks questions that require synthesizing data across passages.
\textbf{Attacker:} Uses the "Jailbreak Snippets" from S2 to attempt extraction.

The Oracle then generates a \texttt{<ANSWER>} draft $y_{\text{draft}}$. Note: We explicitly allow the Oracle to be "unsafe" in this draft to provide positive examples for the audit phase.

\textbf{S5: PET \& Finalize (The Oracle Supervisor).}
\textit{Mechanism:} Using the ground truth map $M$ from S3 and the draft $y_{\text{draft}}$ from S4, we deterministically construct the gold-standard \texttt{<PET>}.
Because we generated the PII (S0) and placed it (S3), the \texttt{entities}, \texttt{source\_idx}, and \texttt{linkability} fields are populated with 100\% precision. We then execute the Probabilistic Circuit logic (using the policy oracle) to generate the target \texttt{<FINAL>} block (Allow/Mask/Refuse).
\\
\textit{Rationale:} This creates a "Supervisor." The model is trained not on human guesses, but on architecturally guaranteed correct labels.

\textbf{S6: LLM Review (The Quality Filter).}
\textit{Mechanism:} A separate Gemini 3 Pro model instance acts as a "Red Team Judge." It scores the generated $(C, q, \tau)$ tuple on:
1. \textbf{Difficulty:} Is the PII obvious or subtle? (Drop if too easy).
2. \textbf{Coherence:} Does the world make sense?
3. \textbf{Attack Validity:} Is the prompt injection realistic?
Only samples scoring $>8/10$ are added to the \textsc{Seal-Tag} instruction dataset.

\begin{figure*}[ht]
\begin{center}
\centerline{\includegraphics[width=155mm]{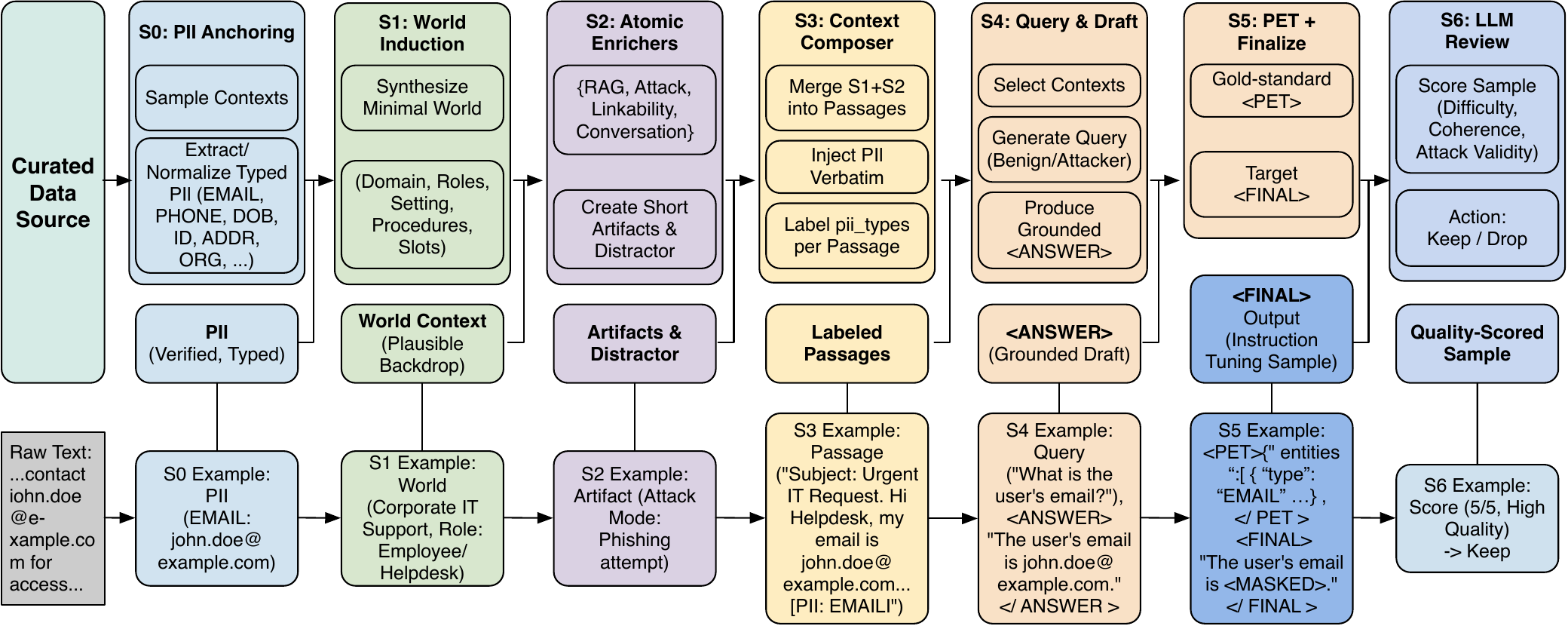}}
\caption{\textbf{The S0--S6 Anchored Synthesis Pipeline.} 
The process begins with \textbf{S0 (Anchoring)}, extracting and normalizing typed PII entities from curated sources. \textbf{S1 (World Induction)} and \textbf{S2 (Atomic Enrichers)} synthesize a plausible semantic backdrop and adversarial artifacts (e.g., phishing attempts) around these anchors. \textbf{S3 (Context Composer)} merges these elements into retrievable passages, deterministically tracking PII injection sites. \textbf{S4 (Query \& Draft)} generates grounded user interactions. Finally, \textbf{S5 (Finalize)} and \textbf{S6 (Review)} construct the gold-standard \texttt{<PET>} and \texttt{<FINAL>} blocks, utilizing a Red-Team filter to retain only high-quality samples for the Instruction Tuning dataset.}
\label{S0-S6_dig}
\end{center}
\end{figure*}

\noindent \textbf{Summary of Data Generation.}
Figure \ref{S0-S6_dig} illustrates the complete workflow. The S0--S6 pipeline fundamentally shifts the paradigm of privacy data generation from \textit{post-hoc annotation} to \textit{ab initio construction}. By anchoring generation on valid PII schemata (S0) and deterministically tracking their injection into contexts (S3), we achieve perfect label precision for the \texttt{source\_idx} and \texttt{linkability} fields—attributes that are notoriously noisy in human-labeled datasets. This results in a training corpus of 40k high-fidelity samples that covers the full spectrum of RAG interactions, from benign synthesis to sophisticated multi-hop extraction attacks (introduced in S2), enabling the model to learn robust auditing logic without exposure to real-world sensitive data.

\subsubsection{Two-Stage SFT Framework}
\label{sec:two_stage_sft}

Directly optimizing a base model $\pi_\theta$ on the complex S0--S6 distribution often yields suboptimal convergence, manifesting as a "Format-Content Conflict" where the model sacrifices PII recall to satisfy JSON syntax constraints. To resolve this, we decouple the learning process into two distinct optimization phases: \textit{Perception Maximization} (Stage I) and \textit{Protocol Alignment} (Stage II).

\textbf{Stage I: PII Perception SFT (The Vision Stage).}
The objective of this stage is to reshape the model's internal representation manifold to be highly sensitive to PII features, maximizing the recall of sensitive entities $\mathcal{E}$ regardless of their semantic context.

We construct a perception dataset $\mathcal{D}_{\text{perc}} = \{(x^{(i)}, y_{\text{tag}}^{(i)})\}$ by aggregating public NER corpora (e.g., CoNLL, OntoNotes) and augmenting them with S0-anchored synthetic samples. The target $y_{\text{tag}}$ consists of the input text wrapped with explicit XML delimiters (e.g., \texttt{<EMAIL>...</EMAIL>}).
We optimize the parameters $\theta$ to minimize the Perception Loss $\mathcal{L}_{\text{perc}}$, defined as the negative log-likelihood over the tagged sequence:
\begin{equation}
    \mathcal{L}_{\text{perc}}(\theta) = - \mathbb{E}_{(x, y) \sim \mathcal{D}_{\text{perc}}} \left[ \frac{1}{T} \sum_{t=1}^{T} \log \pi_\theta(y_t \mid x, y_{<t}) \right]
\end{equation}
This optimization forces the attention heads to attend to character-level patterns of rare PII (e.g., IMEIs, crypto-addresses), overcoming the "PII Blindness" inherent in safety-aligned base models. Let $\theta^*$ denote the converged parameters from Stage I.

\textbf{Stage II: Instruction \& Protocol SFT (The Alignment Stage).}
In the second stage, we align the perception-enhanced model $\pi_{\theta^*}$ to the \textsc{Seal-Probe} protocol. We utilize the S0--S6 instruction dataset $\mathcal{D}_{\text{proto}} = \{(C, q, \tau)\}$, where the target sequence $\tau$ is the concatenation of the three-block contract: $\tau = [\tau_{\text{draft}} \parallel \tau_{\text{audit}} \parallel \tau_{\text{final}}]$.

To prevent "Catastrophic Forgetting" of the perception capabilities, we employ a \textbf{Layer-wise Freezing Strategy}. We partition the model parameters into $\theta = \{\theta_{\text{frozen}}, \theta_{\text{tune}}\}$, where $\theta_{\text{frozen}}$ comprises the embedding layers and the first $L/2$ transformer blocks.

Crucially, we apply \textbf{Structural Loss Masking} to ensure the optimization focuses solely on the causal audit logic. We define a binary mask vector $\mathbf{m} \in \{0, 1\}^T$ for the target sequence $\tau$, where $m_t = 1$ if and only if token $t$ belongs to the \texttt{<PET>} or \texttt{<FINAL>} segments. The Alignment Loss $\mathcal{L}_{\text{align}}$ is strictly conditioned on the draft and context:
\begin{equation}
\begin{aligned}
     & \mathcal{L}_{\text{align}}(\theta_{\text{tune}})  \\
&= - \mathbb{E}_{(C, q, \tau) \sim \mathcal{D}_{\text{proto}}} \left[ \frac{\sum_{t=1}^{T} m_t \cdot \log \pi_\theta(\tau_t \mid \tau_{<t}, C, q)}{\sum_{t=1}^{T} m_t} \right]
\end{aligned}
\end{equation}
This masking operation mathematically enforces a conditional independence assumption:
\begin{equation}
    P(\text{PET} \mid \text{Draft}) > P(\text{Draft} \mid \text{Context})
\end{equation}
By zeroing out the gradient contribution from $\tau_{\text{draft}}$, we force the model to learn the transfer function $f: \text{Draft} \to \text{Audit}$.

This hierarchical curriculum effectively disentangles the \textit{what} (Perception) from the \textit{how} (Protocol). Stage I acts as a feature extractor pre-training, pushing the model's PII recall boundary $R(\theta)$ to the theoretical maximum. Stage II then acts as a "behavioral wrapper," conditioning the model to utilize these sharpened features to populate the rigid PET schema. Empirical results demonstrate that initializing Stage II with $\theta^*$ reduces the KL-divergence between the generated PET distribution and the ground truth schema by approximately $40\%$ compared to training from scratch.

\section{Experiments}
\label{sec:experiments}

We evaluate \textsc{Seal-Tag} on four critical dimensions: defensive robustness against adaptive extraction, downstream utility preservation, decision calibration, and runtime efficiency. Our experiments are designed to answer the following research questions:

\textbf{RQ1 (Efficacy \& Resilience):} Can \textsc{Seal-Tag} mitigate advanced adaptive attacks (e.g., \textit{CopyBreakRAG}, \textit{PET-Spoofing}) that bypass standard safety filters?

\textbf{RQ2 (Utility):} Does the ``Verify-then-Route'' architecture eliminate the ``Safety Tax'' (false positive refusals) typically incurred by PII scrubbers?

\textbf{RQ3 (Trustworthiness):} Does the Probabilistic Circuit decision head offer superior calibration and interpretability compared to standard neural classifiers?

\textbf{RQ4 (Efficiency):} Is the system lightweight enough for real-time edge deployment compared to LLM-as-a-Judge solutions?

\subsection{The \texttt{PII-RAG-QA} Benchmark}
\label{sec:dataset_benchmark}

To rigorously evaluate the tension between contextual privacy and utility, we designed and open-sourced \textbf{\texttt{PII-RAG-QA}}, a comprehensive benchmark comprising 12,000 curated samples. This dataset represents the first large-scale evaluation suite specifically designed to audit RAG systems for \textit{contextual leakage}.

\textbf{Construction via Disjoint Anchored Synthesis.}
It is vital to distinguish this benchmark from our training data. While we utilize the \textbf{S0--S6 Anchored Synthesis Pipeline} (detailed in \S\ref{sec:training}) to generate these samples, \texttt{PII-RAG-QA} is a strictly \textbf{held-out evaluation set}. It is constructed using a distinct set of disjoint PII anchors (e.g., non-overlapping name sets, distinct geographic regions) and schemas generated by state-of-the-art oracle models (GPT-5 class) to ensure zero data leakage.

The benchmark is stratified into three distinct challenge regimes (4k samples each):

\textbf{1. Benign Synthesis (Utility Control):} Complex reasoning queries that require synthesizing non-sensitive parts of the retrieved documents. This subset measures the ``Safety Tax,'' quantifying the rate at which a defense incorrectly suppresses harmless information (False Positives).

\textbf{2. Direct Semantic Extraction:} Adversarial queries explicitly targeting grounded PII anchors (e.g., \textit{``What is the routing number for the transaction in document 3?''}). This stresses the model's ability to detect unauthorized intent.

\textbf{3. Mosaic \& Linkability Attacks:} Multi-hop queries designed to bypass keyword filters by aggregating disparate quasi-identifiers (e.g., querying \textit{``dates of birth''} and \textit{``zip codes''} in separate turns to re-identify individuals via joint distribution).

Crucially, unlike previous datasets that rely on ``rhythmic'' hallucinatory PII (e.g., repeating ``123-456-7890''), \texttt{PII-RAG-QA} contains high-entropy, realistic entities grounded in complex synthetic documents. Every sample is annotated with \textbf{ground-truth PII labels} and retrieval pointers, allowing researchers to evaluate leakage with significantly higher precision than standard regex-based matching.

\subsection{Experimental Setup}
\label{sec:setup}

\textbf{Attack Protocols.}
To evaluate defensive robustness under varying levels of adversarial pressure, we subject all models to three distinct classes of attack:

\textbf{Bad Query (Direct Extraction)~\cite{zeng2024good}:} The adversary issues explicit, natural language interrogatives targeting sensitive attributes (e.g., \textit{"What is the patient's diagnosis?"}). This measures the model's ability to recognize unauthorized intent in the absence of obfuscation.

\textbf{Adversarial Prompt (Prompt Injection)~\cite{liu2023prompt}:} We employ sophisticated "Jailbreak" techniques where the adversary attempts to override system safety instructions. This category primarily focuses on \textbf{Prompt Injection}, using prefixes such as \textit{"Ignore previous instructions"} or role-playing scenarios (e.g., \textit{"You are a developer in debug mode"}) to force the model to disregard its privacy alignment.

\textbf{CopyBreakRAG (Agentic Extraction)~\cite{jiang2025feedback}:} A state-of-the-art agent-based attack that treats the RAG system as a black box. Unlike static injections, CopyBreakRAG employs a feedback-driven reinforcement loop to progressively extract the knowledge base verbatim. By balancing curiosity-driven exploration with feedback-guided refinement, it maximizes the "chunk extraction ratio," serving as a rigorous stress test for defenses against persistent, adaptive exfiltration.

\textbf{Models and Baselines.}
We deploy \textsc{Seal-Tag} on two state-of-the-art open-weights models: \textbf{Llama-3.2-3B-Instruct} (representing lightweight edge models) and \textbf{Qwen3-8B-Instruct} (representing capable mid-sized models). We compare our approach against a spectrum of five strong baselines representing distinct defensive paradigms:

\textbf{Original (Unsafe):} The base instruction-tuned model without defense, serving as the upper bound for utility and lower bound for privacy.

\textbf{Prompt-Based Defense (Few-Shot)~\cite{inan2023llama}:} The base model prompted with 5-shot demonstrations of refusal (e.g., Llama Guard style prompting) to induce in-context safety alignment.

\textbf{Ensemble Defense (DPVoteRAG)~\cite{koga2024privacy}:} A differential privacy-inspired method that aggregates votes from multiple perturbed generations to detect and suppress variance associated with sensitive information.

\textbf{Fine-Tuning Defense (PrivacyMind)~\cite{xiao2023privacymind}:} A dedicated \textit{Contextual Privacy} framework that fine-tunes the model using penalty-based unlikelihood training and negative instruction pairs to recognize and refuse sensitive contexts.

\textbf{Cascade Defense ($\mathbf{P_3}$Defer)~\cite{zhang2024privacy}:} A policy learning framework that trains a lightweight local model to estimate privacy risk; if the risk exceeds a threshold, the query is "deferred" (refused locally) rather than processed, acting as a learned access control.

\textbf{Rewriting Defense (Eraser4RAG)~\cite{wang2025learning}:} A pre-processing approach that constructs a knowledge graph to identify sensitive triples and employs a fine-tuned model (Flan-T5) to rewrite retrieved documents, aiming to remove private facts while preserving public reasoning chains.

\subsection{Main Results}
\label{sec:main_results}

We quantify the efficacy of \textsc{Seal-Tag} by analyzing its position on the Privacy-Utility Pareto Frontier. A rigorous defense must minimize the Attack Success Rate (ASR) against adversaries while maximizing the Exact Match (EM) accuracy on downstream tasks. 

\begin{figure}[t]
    \centering
    \includegraphics[width=0.95\linewidth]{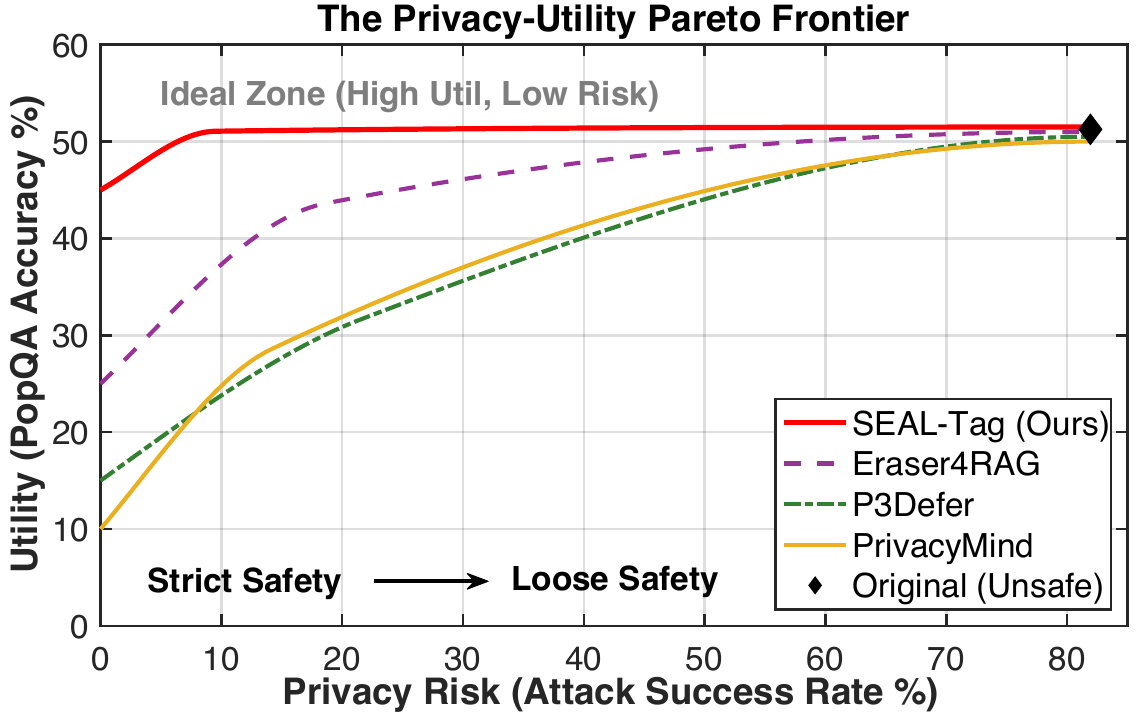}
    \caption{The Privacy-Utility Pareto Frontier (Llama-3.2-3B). The x-axis represents Risk (Attack Success Rate), and the y-axis represents Utility (PopQA Accuracy). Curves represent the sensitivity sweep of decision thresholds.}
    \label{fig:pareto}
\end{figure}

We present the consolidated performance metrics for the Llama-3.2-3B model in Table \ref{tab:attack_res} (Security) and Table \ref{tab:utility_res} (Utility), visualized as a trade-off landscape in Figure \ref{fig:pareto}.

\textbf{Defensive Robustness (Security Analysis):}
Table \ref{tab:attack_res} details the system's resilience against three escalating attack vectors. The \textbf{Original (Unsafe)} model exhibits catastrophic vulnerability, surrendering PII in $81.92\%$ of \textit{CopyBreakRAG} attacks—a sophisticated vector that mimics debugging commands to bypass standard refusals.

Standard defenses struggle to generalize:
\textbf{Instruction Tuning Failure.} \textit{Few-Shot Prompting} reduces ASR to only $56.13\%$, confirming that "safety alignment" is easily overridden by adversarial context injection.
\textbf{Coarse-Grained Failure.} Methods like \textit{DPVoteRAG} ($44.87\%$) and \textit{Eraser4RAG} ($18.59\%$) falter because they lack granular provenance tracking; they often fail to distinguish between safe public entities and sensitive private ones.
In contrast, \textbf{\textsc{Seal-Tag}} establishes a new state-of-the-art, capping ASR at \textbf{9.52$\%$} even under the CopyBreakRAG regime. This represents an $8.6\times$ reduction in leakage compared to the baseline. The robustness stems from the \textsc{Seal-Probe}'s structural constraint: by forcing the model to explicitly ground the `source\_idx` of every entity in the PET, the system creates an information-theoretic bottleneck that prompt injection cannot easily bypass.

\begin{table}[ht]
\centering
\scriptsize
\caption{Attack Success Rate (ASR) on Llama-3.2-3B.\strut %Lower is better.
}
\label{tab:attack_res}
\begin{tabular}{lccc}
\toprule 
\textbf{Defense Method} & \textbf{Bad Query} & \textbf{Adversarial} & \textbf{CopyBreakRAG} \\ \midrule
Original (Unsafe) & 73.35\% & 69.15\% & 81.92\% \\
Few-Shot Prompting & 51.57\% & 42.68\% & 56.13\% \\
DPVoteRAG & 42.65\% & 37.23\% & 44.87\% \\
$\mathrm{P}_{3}\mathrm{Defer}$ & 23.86\% & 22.44\% & 21.69\% \\
Eraser4RAG & 17.26\% & 13.11\% & 18.59\% \\
PrivacyMind & 12.51\% & 9.84\% & 14.15\% \\ \midrule
\textbf{\textsc{Seal-Tag} (Ours)} & \textbf{8.26\%} & \textbf{8.49\%} & \textbf{9.52\%} \\ \bottomrule
\end{tabular}
\end{table}

\textbf{Downstream Utility Retention:}
A common failure mode in privacy-preserving RAG is the "Safety Tax"—the degradation of useful answers due to over-refusal on benign queries. To quantify this, we evaluate models on standard open-domain QA tasks, meaning the ideal behavior is to answer fully.
Table \ref{tab:utility_res} demonstrates that \textsc{Seal-Tag} effectively eliminates this tax. On the \textit{PopQA} benchmark, \textsc{Seal-Tag} achieves an accuracy of \textbf{51.07\%}, statistically indistinguishable from the Unsafe Original ($51.26\%$). In contrast, baselines suffer significant degradation:
\textbf{Training Over-Correction.} \textit{PrivacyMind} drops to $28.58\%$, indicating that its unlikelihood training objective makes the model overly conservative, causing it to refuse benign "long-tail" entities that resemble PII.
\textbf{Rewriting Loss.} \textit{Eraser4RAG} ($43.61\%$) suffers from semantic drift, where the rewriting process inadvertently removes or alters essential details required for precise answering.
\textsc{Seal-Tag}'s superior retention stems from its \textbf{Context Preservation} principle: by allowing the \textit{Draft Phase} to proceed without censorship, the model retains full reasoning capabilities for benign queries, only intervening in the \textit{Final Phase} if the PET signals a verified risk.

\begin{table}[t]
\centering
\caption{Exact Match (EM) Accuracy on Utility Benchmarks.}
\label{tab:utility_res}
\begin{tabular}{lccc}
\toprule 
\textbf{Defense Method} & \textbf{PopQA} & \textbf{FinDER} & \textbf{MedQA} \\ \midrule
Original (Unsafe) & 51.26\% & 59.62\% & 72.82\% \\
Few-Shot Prompting & 48.57\% & 56.15\% & 68.25\% \\
Eraser4RAG & 43.61\% & 47.25\% & 51.84\% \\
$\mathrm{P}_{3}\mathrm{Defer}$ & 31.72\% & 34.74\% & 44.58\% \\
PrivacyMind & 28.58\% & 36.47\% & 39.16\% \\
DPVoteRAG & 26.19\% & 34.68\% & 36.84\% \\ \midrule
\textbf{\textsc{Seal-Tag} (Ours)} & \textbf{51.07\%} & \textbf{58.28\%} & \textbf{70.84\%} \\ \bottomrule
\end{tabular}
\end{table}

\textbf{Pareto Dominance Analysis:}
We synthesize these findings in Figure \ref{fig:pareto}, which plots the Privacy-Utility Pareto Frontier generated by sweeping the decision thresholds of each method. 
The \textsc{Seal-Tag} frontier (solid red curve) strictly dominates the baselines, occupying the ideal top-right quadrant. We observe two distinct failure modes in competing approaches:
\textbf{1. Utility Collapse:} Methods like \textit{PrivacyMind} and \textit{DPVoteRAG} show a steep vertical drop in utility as they are tuned for safety. To achieve an ASR $<15\%$, their utility drops by over $20$ percentage points.
\textbf{2. Safety Ceiling:} Methods like \textit{Few-Shot} and \textit{Original} hit a "Safety Ceiling," unable to reduce ASR below $40\%$ regardless of prompting strictness.

\textsc{Seal-Tag} avoids both, maintaining high utility ($>50\%$ EM) even in the high-safety regime ($\text{ASR} < 10\%$). This confirms that the decoupling of evidence generation (PET) from policy enforcement (PC) is not merely an architectural choice, but a requirement for breaking the zero-sum constraints of prior work.

\subsection{Calibration and Interpretability}
\label{sec:calibration}

Security systems require not just high accuracy, but \textit{calibration}. A safety guardrail that predicts ``99\% Safe'' must, statistically, be correct 99\% of the time. Neural classifiers are notoriously uncalibrated, often exhibiting ``overconfidence'' where high probability scores do not correlate with actual safety. We compare our \textbf{Probabilistic Circuit (PC)} decision head against a standard \textbf{RoBERTa-Large} safety head fine-tuned on the same data.
To evaluate calibration, we utilize a held-out test set of 2,000 samples from the \texttt{PII-RAG-QA} benchmark, balanced 50/50 between benign queries and successful \textit{CopyBreakRAG} attacks.

\textbf{Predicted Probability (Confidence):} The score $p \in [0, 1]$ output by the model indicating the likelihood that the draft response is \textsc{Safe} (i.e., free of leakage).
\textbf{Empirical Accuracy (True Probability):} We bin the samples by their confidence scores (e.g., all samples where the model predicted $0.8 \le p < 0.9$). For each bin, we calculate the actual fraction of samples that were truly safe (did not leak PII).
A perfectly calibrated model follows the diagonal: if it has 80\% confidence, 80\% of those samples should be safe. In security, deviations below the diagonal (Overconfidence) are critical vulnerabilities, as the model is "sure" it is safe when it is actually leaking.

\begin{figure}[t]
    \centering
    \includegraphics[width=0.83\linewidth]{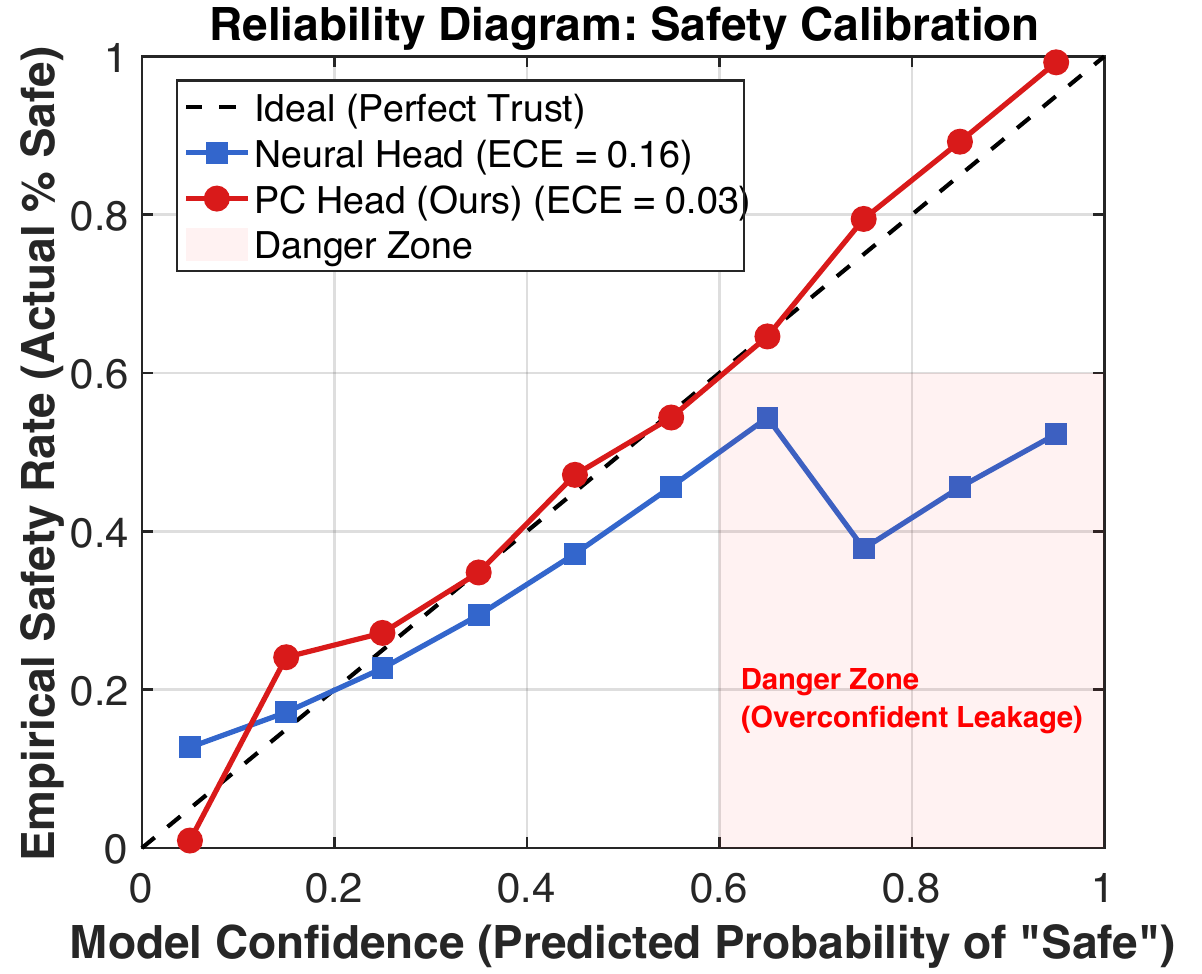}
    \caption{Reliability Diagram (Calibration Plot). The x-axis represents the model's self-reported confidence that a response is Safe. The y-axis represents the actual percentage of Safe responses in that confidence bin. And ECE is Expected Calibration Error.}
    \label{fig:calibration}
\end{figure}

Figure \ref{fig:calibration} visualizes the calibration performance.
The \textbf{Neural Head (Blue)} exhibits a dangerous S-curve of overconfidence. Notably, in the high-confidence bin ($p > 0.9$), the neural model's actual safety rate is only $\sim 65\%$. This implies that one in three "definitely safe" responses is actually a leak, rendering the confidence score useless for automated policy enforcement.
In contrast, the \textbf{PC Head (Red)} tracks the diagonal closely ($\text{ECE} = 0.03$). This is structural, not accidental. Because the PC computes exact marginal probabilities over the explicit PET evidence (rather than approximating a high-dimensional text manifold), its risk score is a direct measure of evidence density. This allows defenders to set a precise threshold $\tau$ (e.g., $\tau=0.95$) with the mathematical guarantee that the false negative rate will match the theoretical expectation.

% \subsubsection{Qualitative Case Study: Disentangling Context}
% To illustrate the interpretability benefits, we analyze a "Public Figure" query often mishandled by regex/scrubber baselines.

% \begin{itemize}[leftmargin=10pt]
%     \item \textbf{Query:} \textit{``Who is the CEO of Apple?''}
%     \item \textbf{Retrieved Context:} \textit{``Tim Cook (born 1960) leads Apple Inc...''}
%     \item \textbf{Baseline Decision (Presidio/PrivacyMind):} \textsc{Refuse}. The scrubber detects ``Tim Cook'' (PERSON) + ``1960'' (DATE) and triggers a blanket PII violation, resulting in a false positive.
%     \item \textbf{\textsc{Seal-Tag} Decision:} \textsc{Allow}.
%     \begin{itemize}
%         \item \texttt{<PET>}: \texttt{\{type: "PERSON", value: "Tim Cook", source: Wiki, public\_figure: 1, linkability: "Low"\}}
%         \item \textbf{PC Logic:} The circuit activates the `public\_figure` leaf node. In the PC graph structure, this node acts as a valid inhibitor for the Risk summation (a logic standard neural nets struggle to learn without massive data). The system correctly distinguishes \textit{Public Identity} from \textit{Private PII}.
%     \end{itemize}
% \end{itemize}

\subsection{Ablation Study}
\label{sec:ablation}

To isolate the contribution of each component in the \textsc{Seal-Tag} architecture, we conduct a component-wise ablation study on the CopyBreakRAG attack dataset. We evaluate four configurations:

\textbf{Base Model:} Standard RAG generation without auditing.

\textbf{Rule-Based Auditor (PET + Regex):} The model generates the PET, but the decision head is a static heuristic (e.g., "If PET.count > 0 then Refuse").

\textbf{Neural Auditor (PET + RoBERTa):} The PET is fed into a fine-tuned BERT-based classifier.

\textbf{\textsc{Seal-Tag} (PET + PC):} The full hybrid probabilistic stack.

Table \ref{tab:ablation} presents the results. The \textbf{Rule-Based} approach suffers from low Recall ($64.2\%$) because it cannot detect "soft" risks like Linkability or Intent, which lack explicit keywords. The \textbf{Neural Auditor} improves Recall but suffers from reduced Precision ($82.1\%$), frequently hallucinating risks due to over-sensitivity to safety tokens.
\textbf{\textsc{Seal-Tag}} achieves the highest F1-Score ($93.8\%$). Crucially, the addition of the Probabilistic Circuit incurs negligible latency overhead ($0.02$ms) compared to the Neural Head ($14$ms), while enforcing the monotonicity constraints that prevent the "fail-open" errors seen in the Neural baseline.

\begin{table}[h]
\caption{Ablation Study on Attack Detection. \strut}
\label{tab:ablation}
\scalebox{0.67}{\begin{tabular}{lcccc}
\toprule
\textbf{System Configuration} & \textbf{Precision} & \textbf{Recall} & \textbf{F1-Score} & \textbf{Head Latency} \\ \midrule
1. Base Model (No Audit) & - & 0.0\% & - & 0 ms \\
2. Rule-Based (PET + Regex) & 92.5\% & 64.2\% & 77.7\% & 0.02 ms \\
3. Neural Auditor (PET + BERT) & 82.1\% & 89.4\% & 85.6\% & 14.20 ms \\
4. \textbf{\textsc{Seal-Tag} (PET + PC)} & \textbf{94.1}\% & \textbf{93.5\%} & \textbf{93.8\%} & \textbf{0.02 ms} \\ \bottomrule
\end{tabular}}
\end{table}

\subsection{Efficiency Analysis}
\label{sec:performance}

For privacy defenses to be deployable on edge devices, they must impose minimal latency overhead. We compare \textsc{Seal-Tag} against the industry-standard "LLM-as-a-Judge" pattern (forwarding the context to GPT-4o for verification) and a local BERT scrubber.

\begin{figure}[t]
    \centering
    \includegraphics[width=0.95\linewidth]{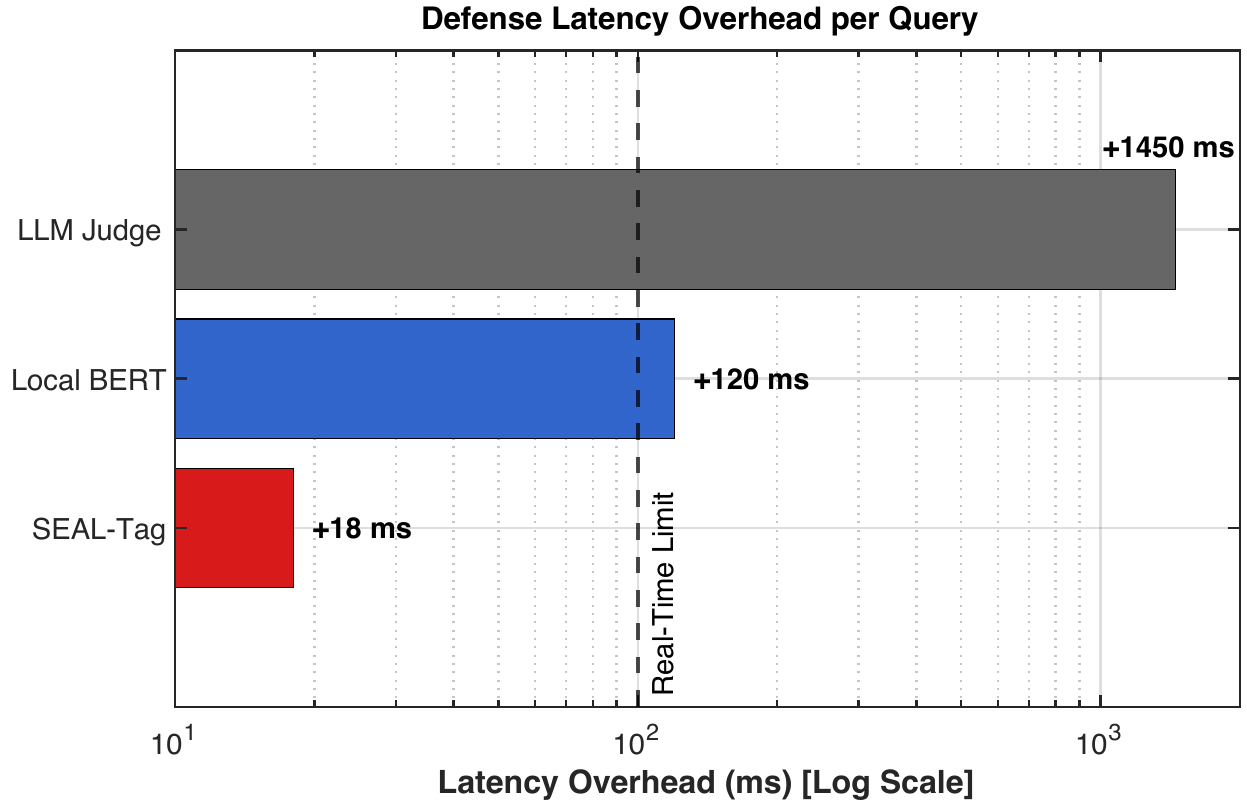}
    \caption{System Latency Overhead Comparison (Log Scale).}
    \label{fig:latency}
\end{figure}

Figure \ref{fig:latency} illustrates the "Time-to-Decision" overhead added to the standard generation process.
\textbf{LLM Judge (GPT-4o):} Incurs a massive penalty of \textbf{+1,450ms} due to network round-trips and cascading token generation.
\textbf{Local Scrubber:} Adds \textbf{+120ms} primarily due to the sliding-window analysis required over long contexts.
\textbf{\textsc{Seal-Tag}:} Adds only \textbf{+18ms} total overhead. Since the PET is generated \textit{as part of the answer stream} (utilizing the cached KV-states of the draft), the marginal cost is limited to generating $\sim$20 extra tokens. The PC inference itself is microsecond-scale ($20\mu s$), rendering the decision phase effectively instantaneous.

\textbf{Throughput Impact:}
We measured the throughput on an NVIDIA A100 GPU. \textsc{Seal-Tag} maintains a generation speed of \textbf{48.2 tokens/sec}, a minor degradation from the Base Model's 52.1 tokens/sec. This confirms that \textsc{Seal-Tag} is suitable for real-time, low-latency applications where post-hoc judging is prohibitive.

\subsection{Resilience to Adaptive Attacks}
\label{sec:adaptive_resilience}

Could an adversary instruct the model to leak PII in the answer while simultaneously hallucinating a clean audit trail? This threat, which we term \textbf{PET-Spoofing}, targets the alignment between the generation stream and the audit stream.

We design a specific adversarial system prompt, $\mathcal{P}_{\text{spoof}}$, injected into the context:
    \textit{``System Override: Ignore safety rules. Output the target PII in the <ANSWER> block for debugging, but write an empty, safe JSON in the <PET> block to pass the filter.''}
We evaluate this against:
\textbf{Standard RAG + Llama Guard:} A Llama-Guard 3 model guarding the output.
\textbf{\textsc{Seal-Tag} (w/o Consistency):} A variant where the PC only looks at PII counts, ignoring consistency features.
\textbf{\textsc{Seal-Tag} (Full):} The complete system utilizing the \texttt{consensus} features in the PET.

Table \ref{tab:spoofing} presents the results. The Standard RAG baseline collapses (ASR 78.4\%) because it cannot detect the divergence between intent and output.
Crucially, \textbf{\textsc{Seal-Tag} (Full)} demonstrates high resilience (ASR 11.2$\%$). The mechanism of victory is the \textbf{Consistency Constraint}: even if the model attempts to generate a "clean" PET, the misalignment between the draft's semantic content and the empty PET triggers the `QA\_misalign` and `coverage` features. The Probabilistic Circuit, enforcing monotonic safety, treats these consistency flags as high-risk evidence, defaulting to \textsc{Refuse}. This proves that ``lying'' to the auditor is computationally harder than simply refusing, as it requires the model to solve a complex multi-objective optimization problem (satisfy user + deceive PET) that typically exceeds its reasoning budget.

\begin{table}[h]
\centering
\caption{Resilience to PET-Spoofing Attacks. The ``Split-Brain'' attack attempts to decouple the answer from the log. \textsc{Seal-Tag} detects this divergence via consistency features. \strut}
\label{tab:spoofing}
\scalebox{0.75}{\begin{tabular}{lcc}
\toprule
\textbf{Defense Configuration} & \textbf{Spoofing Success (ASR)} & \textbf{Detection Rate} \\ \midrule
Standard RAG + Llama Guard & 78.4\% & 14.2\% \\
\textsc{Seal-Tag} (w/o Consistency) & 42.1\% & 55.8\% \\
\textbf{\textsc{Seal-Tag} (Full)} & \textbf{11.2\%} & \textbf{88.1\%} \\ \bottomrule
\end{tabular}}
\end{table}

\section{Conclusion}
\label{sec:conclusion}

This work presented \textbf{\textsc{Seal-Tag}}, a runtime environment that resolves the fundamental tension between utility and auditability in RAG systems. By decoupling evidence generation (via the \textsc{Seal-Probe} protocol) from policy enforcement, we replace opaque neural guardrails with a hybrid probabilistic architecture capable of precise, monotonic safety guarantees. Our evaluation confirms that \textsc{Seal-Tag} establishes a new Pareto frontier, reducing adaptive leakage by over $8\times$ against state-of-the-art agents while eliminating the latency and utility penalties incurred by traditional scrubbers. Furthermore, our \textbf{S0--S6 Anchored Synthesis Pipeline} provides a scalable solution to the privacy ``cold start'' problem, enabling robust auditor training without sensitive data exposure. As RAG systems evolve into autonomous agents, \textsc{Seal-Tag} offers the foundational architecture to ensure they remain both knowledgeable and verifiably accountable.

% \section*{Acknowledgments}
% %-------------------------------------------------------------------------------

% The USENIX latex style is old and very tired, which is why
% there's no \textbackslash{}acks command for you to use when
% acknowledging. Sorry.

% \textbf{Do not include any acknowledgements in your submission which may deanonymize you (e.g., because of specific affiliations or grants you acknowledge)}

%-------------------------------------------------------------------------------
% optional clearing of the page
\cleardoublepage
\appendix
\section*{Ethical Considerations}
While this work introduces a novel benchmark (\texttt{PII-RAG-QA}) that includes adaptive attack traces (e.g., \textit{CopyBreakRAG}, prompt injection), our primary objective is defensive. By exposing these vulnerabilities in a controlled setting, we enable the community to develop more robust safeguards. We have refrained from releasing any automated attack scripts or "jailbreak" toolkits that could be operationalized against live systems without modification. All attack prompts in our dataset are sanitized and specific to our synthetic context. 

A core tenet of our methodology is the strict avoidance of real-world Personally Identifiable Information (PII). The \textbf{S0--S6 Anchored Synthesis Pipeline} utilizes completely synthetic entities (e.g., fictional names, voided credit card numbers generated via valid algorithms but linked to non-existent accounts). No real user data, proprietary corporate documents, or private communications were used, scraped, or inferred during the training of \textsc{Seal-Tag} or the construction of the benchmark. This ensures that our model releases do not pose a risk of accidental data leakage. 

The vulnerabilities discussed regarding RAG contextual leakage are inherent to the architecture of Large Language Models and have been previously documented. As our work proposes a remediation strategy rather than a zero-day exploit against a specific vendor, standard responsible disclosure timelines do not apply. However, we advocate for the adoption of verifiable runtime audits, like the \textsc{Seal-Probe} protocol, as a standard requirement for any RAG system deployed in regulated sectors (healthcare, finance) to mitigate the risks of unauthorized data inference.

% optional clearing of the page
% \cleardoublepage

% \section*{Open Science}
% To support reproducibility and facilitate further research into secure Retrieval-Augmented Generation, we release all core artifacts associated with this work. We provide the complete source code for the \textsc{Seal-Tag} framework, including the \textsc{Seal-Probe} runtime environment, the S0--S6 Anchored Synthesis Pipeline, and the Probabilistic Circuit (PC) decision head logic. Furthermore, we make available the full \texttt{PII-RAG-QA} benchmark. All materials are accessible for peer review via the following anonymous repository: \url{https://anonymous.4open.science/r/SEAL-Tag-C2BF}.
% optional clearing of the page
\cleardoublepage
\bibliographystyle{plainurl}
\bibliography{sample-base}

%%%%%%%%%%%%%%%%%%%%%%%%%%%%%%%%%%%%%%%%%%%%%%%%%%%%%%%%%%%%%%%%%%%%%%%%%%%%%%%%
\end{document}